# Evaluation of the potential of Near Infrared Hyperspectral Imaging for monitoring the invasive brown marmorated stink bug

Veronica Ferrari [a], Rosalba Calvini [a,*], Bas Boom [b], Camilla Menozzi [a], Aravind Krishnaswamy Rangarajan [b], Lara Maistrello [a], Peter Offermans [b], Alessandro Ulrici [a]

[a] Department of Life Sciences, University of Modena and Reggio Emilia, Pad. Besta, Via Amendola, 2, 42122, Reggio Emilia, Italy

[b] IMEC OnePlanet, Bronland 10, Wageningen, The Netherlands

* Corresponding author: rosalba.calvini@unimore.it

## Abstract

The brown marmorated stink bug (BMSB), *Halyomorpha halys*, is an invasive insect pest of global importance that damages several crops, compromising agri-food production. Field monitoring procedures are fundamental to perform risk assessment operations, in order to promptly face crop infestations and avoid economical losses. To improve pest management, spectral cameras mounted on Unmanned Aerial Vehicles (UAVs) and other Internet of Things (IoT) devices, such as smart traps or unmanned ground vehicles, could be used as an innovative technology allowing fast, efficient and real-time monitoring of insect infestations.

The present study consists in a preliminary evaluation at the laboratory level of Near Infrared Hyperspectral Imaging (NIR-HSI) as a possible technology to detect BMSB specimens on different vegetal backgrounds, overcoming the problem of BMSB mimicry. Hyperspectral images of BMSB were acquired in the 980-1660 nm range, considering different vegetal backgrounds selected to mimic a real field application scene. Classification models were obtained following two different chemometric approaches. The first approach was focused on modelling spectral information and selecting relevant spectral regions for discrimination by means of sparse-based variable selection coupled with Soft Partial Least Squares Discriminant Analysis (s-Soft PLS-DA) classification algorithm. The second approach was based on modelling spatial and spectral features contained in the hyperspectral images using Convolutional Neural Networks (CNN). Finally, to further improve BMSB detection ability, the two strategies were merged, considering only the spectral regions selected by s-Soft PLS-DA for CNN modelling.

# 1 Introduction

In the last decades, the increase of anthropogenic activities determined the spread of invasive insect pests, which can seriously affect agroecosystems compromising agri-food production and resulting in severe economic losses. One of the most worrisome pests of global importance is the brown marmorated stink bug (BMSB), *Halyomorpha halys,* which causes serious damages to several agricultural crops (Leskey and Nielsen, 2018). The damage occurs mainly to fruits and seeds as a result of BMSB feeding activity: its piercing-sucking mouth apparatus determines deformities, scars, discolorations and pitting which make products unmarketable (Rice et al., 2014). In Southern Europe, suitable climate and high density of crops provided excellent conditions for the establishment of large populations of BMSB (Costi et al., 2017). As a result of BMSB activity, in 2019 economical losses in fruit orchards of Northern Italy were estimated to be equal to €590 million (CSO, 2020).

The management of BMSB is very challenging due to high reproductive potential, high mobility, polyphagy and ineffectiveness of available broad-spectrum insecticides, which determine a negative impact on the environment (Leskey, Lee, et al., 2012; Leskey, Short, et al., 2012; Maistrello et al., 2017, 2018). According to Integrated Pest Management (IPM) practices, field monitoring of insect pests is of fundamental importance to gain information about their presence and to timely adopt proper actions to face the infestation and avoid economical losses. Although it represents a crucial step, field monitoring is time and money consuming for farmers, since it requires direct field inspection by technicians (Maistrello et al., 2017, Rice et al., 2014).

With the aim of improving crop field pest management, automated monitoring systems based on spectral cameras mounted on Unmanned Aerial Vehicles (UAVs) and other Internet of Things (IoT) devices, such as smart traps or unmanned ground vehicles, can be used as an innovative technology allowing fast, efficient, and real-time monitoring (Betti-Sorbelli et al., 2022, Friha et al., 2021; Miella et al., 2019). As previous studies reported, multispectral imaging (MSI) and hyperspectral imaging (HSI) cameras in the visible and near infrared regions mounted over UAVs proved to be effective tools for a fast and reliable assessment of crop infection and infestation (Caballero et al., 2020; Calvini et al., 2020). Consequently, the time needed to detect the presence of possible crop infections is strongly reduced and it is possible to employ targeted pest management strategies.

The present study performed an initial assessment at the laboratory level of hyperspectral imaging as a possible method to identify BMSB specimens on different background types simulating a real field application scene. In this case, the effectiveness of spectral cameras working in the near infrared range (NIR) has been evaluated in order to overcome BMSB mimicry, as the brown marmorated colour makes this insect hardly detectable on dark brown vegetal backgrounds with spectral cameras operating in the visible range.

Hyperspectral images are three-dimensional matrices composed of one spectral ($\lambda$) and two spatial ($x$ and $y$) dimensions, obtained by stacking together hundreds of ($x$, $y$) grey-scale images acquired at different successive wavelengths, $\lambda$. Therefore, this technique couples the advantages of spectroscopic methods with the possibility of visualizing spectral data at each pixel of an image, allowing the visualization of the chemical composition of the sample surface (Gowen et al., 2007, 2011). Despite the great potential of this technique, its data-richness represents at the same time the main advantage and disadvantage of HSI: a large amount of data permits a detailed representation of the analysed samples, but, at the same time, it involves issues related to data handling, storage and analysis.

Chemometric techniques are mandatory to unravel the curse of dimensionality in HSI and to develop classification or calibration models able to predict the qualitative or quantitative properties of interest from hyperspectral data (Saha & Manickavasagan, 2021). Simple but effective applications of chemometric techniques to hyperspectral data include the use of linear calibration or classification methods, such as Partial Least Squares (PLS) or Partial Least Squares Discriminant Analysis (PLS-DA) (Burger & Gowen, 2011; Calvini et al., 2015). An advantage of these methods consists in the fact that the models are easily interpretable, especially when it is necessary not only to obtain good model performances but also to highlight the relevant spectral regions for the problem at hand. In this context, it is also possible to apply spectral variable selection algorithms, which allows the subsequent elimination of spectral regions that are not pertinent, leading at the same time to better results in classification or calibration issues and to an increased chemical interpretation of the results (Calvini et al., 2015; Xiaobo et al., 2010). Considering HSI applications, the identification of spectral bands relevant for the problem at hand allows to further develop faster and cheaper multispectral imaging systems. In MSI systems only a limited number of specific wavebands are considered, reducing the time needed for the acquisition and the efforts for data management. In addition, MSI systems are characterised by higher resistance and stability of the optical components, which makes them more suitable for application in the field (Calvini et al., 2017; Gowen et al., 2007).

Recent progresses in computer technology led to the development of advanced Deep Learning (DL) techniques, which require large datasets to be trained and are able to face complex applications. Among DL methods, Artificial Neural Networks (ANNs) progressively emerged as valuable tools to solve complex and highly non-linear regression and classification tasks, besides becoming popular in countless fields (Brown, Tauler & Walczak, 2020).

In particular, Convolutional Neural Networks (CNNs) achieved state-of-the-art performances in the domain of Computer Vision applications. In this context, CCNs models have been successfully employed for image classification, object detection, and image segmentation (Shorten & Khoshgoftaar, 2019). To optimize CNNs modelling capabilities, multiple architectures have been proposed for image segmentation and classification (Badrinarayanan et al., 2017, Long et al., 2015, Ronneberger et al., 2015) including the U-Net (Ronneberger et al., 2015), which is of particular interest for the implementation on edge devices such as UAVs, due to its ability to achieve good model performances with reduced computation efforts (Liu et al., 2021). U-Net was initially designed for computer vision applications, however recent works also demonstrated the advantages of this algorithm in the segmentation of hyperspectral images (Cervantes et al., 2021, Moustafa et al., 2021). In contrast to PLS-DA, the U-Net employs both spatial and spectral information to perform the classification by virtue of its 2D convolutional layers which are designed to exploit spatial relationships between pixels. Because of this advantage, the U-Net is able to suppress false positive pixels by considering neighbouring pixels values, which should make the U-Net more robust for application in the field. In this study, NIR hyperspectral imaging was used to develop classification models able to discriminate BMSB specimens and the vegetal backgrounds following different Machine Learning (ML) strategies. Firstly, the spectral information was modelled using Soft Partial Least Squares-Discriminant Analysis (Soft PLS-DA), an extension of the classical PLS-DA algorithm, and by s-Soft PLS-DA (Calvini et al., 2018), a version of Soft PLS-DA where sparse-based variable selection was also implemented (Calvini et al., 2015, 2017; Filzmoser et al., 2012). Then, a deep learning method based on U-Net was implemented and adapted for hyperspectral images, focusing on the spatial features of BMSB. Finally, in order to reduce the complexity of the network in the spectral dimension, the U-Net was adjusted to use the relevant spectral regions previously identified by s-Soft PLS-DA. This also allowed us to investigate whether merging together a spectral-based and a spatial-based method could result in improved classification performances.

# 2 Materials and Methods

## 2.1 Samples

BMSB specimens used in the present work were provided by the Applied Entomology Lab, University of Modena and Reggio Emilia. All the progenitor individuals had been previously captured in urban parks of the city of Reggio Emilia using the tree beating technique. The bugs were reared in climatic chambers at 26 °C, 60% relative humidity, L16:D8 photoperiod, inside clear mesh cages (30 × 30 × 30cm, approximately 40 individuals/cage) with organic tomatoes, carrots, green bean pods and raw peanuts as food. A bottle cap with a water-soaked cotton swab was used as water supply. Food and water were replaced twice per week.

The samples of vegetal backgrounds were collected in the University campus surroundings (Via Amendola, 2, San Maurizio, Reggio Emilia) on the same day of image acquisition.

To develop the classification models and to identify the spectral bands able to discriminate BMSB from the different backgrounds, 20 specimens of BMSB and seven types of vegetal backgrounds were considered, corresponding to green leaves, yellow leaves, dry leaves, grass, soil, bark and tree branches. These background types were selected to mimic real field conditions. BMSB specimens were randomly divided in five different groups (G1-G5) with 4 insects in each group; the bugs belonging to each group were kept together and always acquired in the same image of the different vegetal background types.

## 2.2 Image acquisition and elaboration

The hyperspectral images were acquired using a HSI line-scan system composed of a desktop NIR Spectral Scanner (DV Optic) embedding a Specim N17E reflectance imaging spectrometer, coupled to a Xenics XEVA 1.7-320 camera (320 × 256 pixels) embedding Specim Oles 31 f/2.0 optical lens and covering the spectral range from 900 to 1700 nm (5 nm resolution, 150 spectral channels). To enable a better evaluation of the stability of the acquisition system over time, a setup composed of a silicon carbide sandpaper as sample background, which is characterized by a very low and constant reflectance spectrum (Burger & Geladi, 2006), a 99% reflectance standard, and two ceramic tiles with two different grayscale tones and intermediate reflectance values were used for the acquisition of all the images.

The raw data were then converted into reflectance values by applying the instrument calibration procedure based on the high-reflectance standard reference and on the estimate of the dark current (Burger & Geladi, 2005). Furthermore, to minimize the variability among images over time, an additional internal calibration was performed (Ulrici et al., 2013).

The wavelengths at the extremes of the spectral range are characterized by low S/N values and were excluded, considering only the spectral range between 980 and 1660 nm (137 wavelengths) for further analysis.

On the whole, 35 hyperspectral images were acquired (= 7 vegetal backgrounds × 5 BMSB groups). The BMSB specimens belonging to each group were positioned in different image area and considering different orientations of the bugs when acquiring the samples on the different background types.

Firstly, the pixels related to the black sandpaper background were removed from each image by excluding all the pixels with reflectance values lower than 0.3 reflectance units at 1000 nm, which was identified as the most discriminant wavelength between the sample area and the black sandpaper background. Then, Principal Component Analysis (PCA) was applied to each image to perform a masking procedure, in order to separate the pixels belonging to the bugs from the pixels belonging to the different vegetal backgrounds. In this case, the hyperspectral images were preprocessed using standard normal variate (SNV) and mean center. Therefore, for each hyperspectral image two masks were obtained: one to identify the pixels belonging to the vegetal backgrounds and one to identify the pixels belonging to the BMSB specimens.

To perform image elaboration, the acquired hyperspectral images were converted to .mat format and further analysed in MATLAB environment (R2020b, The MathWorks Inc., USA). Image correction based on internal calibration was performed using *ad hoc* routines written in MATLAB environment, while the masking procedure based on PCA was performed using the HYPER-Tools software package (Mobaraki & Amigo, 2018) (version 3.0, https://www.hypertools.org).

## 2.3 Data analysis

In this study the acquired hyperspectral images were used to develop classification models able to discriminate BMSB bugs from vegetal backgrounds following two main strategies. The former approach was focused on modelling the different spectral features of the considered classes using the Soft PLS-DA (Calvini et al., 2018) algorithm. In addition, this kind of approach also allowed to select the more relevant spectral regions by coupling sparse-based variable selection with Soft PLS-DA (Calvini et al., 2015; Filzmoser et al., 2012; Rasmussen & Bro, 2012). The latter approach utilized a U-Net, a Convolutional Neural Network (CNN) based deep learning model, that allows to exploit the features based on spatial relationships between neighbouring pixels of BMSB.

Then, the advantages of the two different strategies were combined to further implement a more effective detection method. As a matter of fact, spectral variable selection enabled the identification of a reduced subset of spectral bands relevant for BMSB detection, which was used to improve the detection ability of the U-Net classification model based on BMSB spatial features (**Figure 1**).

Finally, the performance of all the classification models, that were developed at the pixel-level, was also evaluated at the object-level. In pixel-level classification, each pixel spectrum of an image is considered as a separate entity and it is classified by the model into the corresponding predicted class. Instead, in the case of object-level classification, each BMSB specimen is considered as a single object and classification performance was determined by the correct identification of the considered BMSB specimens.

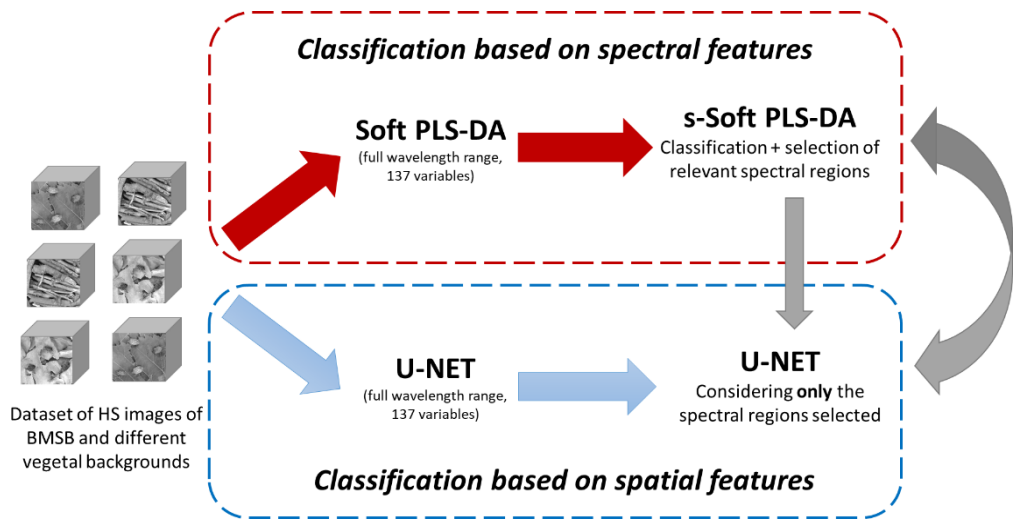

*Figure 1.* Schematic representation of the classification strategies adopted to detect BMSB on different vegetal backgrounds using NIR-HSI.

### 2.3.1 Classification based on spectral features

For the development of the classification models using Soft PLS-DA and s-Soft PLS-DA it is necessary to select a set of representative spectra belonging to both BMSB and vegetal backgrounds. This phase is crucial since it determines the representativity of spectral signatures considered for the two classes to implement robust and reliable classification models.

For each image, a PCA model was calculated using mean center as preprocessing method, considering only the pixels belonging to the vegetal background and retaining 3 PCs. The number of PCs to retain in the model was chosen based on a preliminary evaluation performed on some representative sample images. Then, outlier pixels were removed considering the 99.9% confidence limit for both Hotelling $T^2$ values and Q residuals. Finally, a new PCA model was calculated considering again 3 PCs and the Kennard-Stone algorithm (Kennard & Stone, 1969) was applied in the PCs space to select for each image 200 spectra representative of the vegetal background. As a result, 7000 spectra belonging to the vegetal backgrounds were collected (= 200 spectra × 35 hyperspectral images).

The same procedure was also adopted to select from each image 200 spectra belonging to the bugs, obtaining also in this case 7000 spectra representative of the BMSB class.

Therefore, the dataset of spectra belonging to the modelled classes and used to develop the classification models was composed of 14000 spectra.

The dataset was split into a training set (TR-spectra), used for model calculation, and a test set (TS-spectra) used for external validation. To this aim, BMSB specimen groups (previously cited in **Section 2.1**) were considered: the spectra belonging to images containing G1-G3 groups were used for the TR-spectra dataset, including 8400 spectra (= 3 BMSB groups × 7 backgrounds × 200 spectra × 2 classes), while the spectra belonging to images containing G4 and G5 groups were considered for the TS-spectra dataset, including the remaining 5600 spectra.

As an additional external validation, the classification models were also applied to the test set images (TS-images), i.e., to the whole hyperspectral images of the G4 and G5 BMSB specimen groups. The corresponding prediction images (i.e., the images with the pixels coloured according to the predicted class) were used to visualize the prediction performances directly on the images and to obtain a quantitative evaluation of the classification performances over the whole images.

Therefore, the external validation of the classification models was performed considering images containing BMSB specimens different from those contained in the images used as training set to calculate the classification models.

The classification models were calculated both using Soft PLS-DA and combining this algorithm with a sparse-based variable selection approach.

Soft PLS-DA combines the advantages of classical discriminant analysis and class modelling techniques, in order to increase the flexibility of classification models for field application.

Like PLS-DA, the Soft PLS-DA algorithm is based on a discriminant approach, which maximizes the discrimination between samples belonging to the investigated classes, but, unlike PLS-DA, class assignment is performed by fixing additional limits both on the Y predicted values and on the Q residuals. More in detail, a new sample is assigned to a defined class according to the following criteria:

- having Q residuals values falling inside the 99.9 % confidence limit of the model. This limit has been chosen to set boundaries large enough to consider different classes' variability as much as possible while being able to exclude samples with a very low fit to the model;

- having *y* predicted values falling inside an acceptability range for the considered class, whose lower limit is defined by PLS-DA threshold for the investigated class while the upper limit allows to reject objects found at the extremes of the Gaussian probability density function.
- for classification problems with more than two classes, the samples must be unambiguously assigned only to one class.

Samples that do not match all the three criteria defined by Soft PLS-DA decision rule are not assigned to any class and automatically labelled as "not assigned" samples (NA). In this manner, Soft PLS-DA overcomes PLS-DA limited ability to correctly handle new objects not belonging to the target classes while maximizing discrimination between the classes of interest (Barker & Rayens, 2003; Calvini et al., 2018; Pomerantsev & Rodionova, 2018). For a detailed description of Soft PLS-DA algorithm the reader is referred to Calvini et al. (2018).

Moreover, in s-Soft PLS-DA sparse-based variable selection was coupled with Soft PLS-DA, to maintain high model performances while selecting only the most representative wavelengths involved in the classification. The main idea behind sparse methods in the context of linear regression and classification is to decrease the computational load while increasing the robustness of the prediction models (Burger & Gowen, 2011). The sparsity is achieved by adding a penalty term to the computation of the model coefficients: in this case, a Least absolute shrinkage and selection operator (Lasso) penalty approach was applied (Calvini et al., 2017, 2018; Filzmoser et al., 2012; Tibshirani, 1996).

Both Soft PLS-DA and s-Soft PLS-DA classification models were calculated considering different row-preprocessing methods, i.e., SNV, detrend, first derivative and second derivative, followed by mean center.

The optimization of the classification models was performed using venetian blinds cross-validation with 3 deletion groups.

In s-Soft PLS-DA it is necessary to also optimize the sparsity of the model, i.e., the number of variables to be selected, in addition to the proper number of LVs. Different models were calculated considering all the possible combinations between a number of LVs ranging from 1 to 10 and a number of variables selected for each LV ranging from 5 to 137, with a step equal to 5. The best combination between the number of LVs and the number of selected variables was identified by maximising classification efficiency (EFF, *see* Section 2.3.5) estimated in cross-validation (Calvini et al., 2018)

Soft PLS-DA and s-Soft PLS-DA were calculated using *ad hoc* routines written in MATLAB environment (ver. 2020b, The MathWorks, USA). The MATLAB routine to run Soft PLS-DA algorithm (Calvini et al., 2018) is freely downloadable from
http://www.chimslab.unimore.it/downloads/.

### 2.3.2 Classification based on spatial features

Classification based on spatial features using Deep Learning approaches relies on big data: a large amount of training data is required to maximize the generalizability of the model in order to avoid overfitting (Shorten & Khoshgoftaar, 2019). Therefore, when dealing with smaller datasets it is a common practice to apply Data Augmentation techniques such as rotation, flipping, and scaling to obtain an augmented experimental dataset. In this process, the total amount of available images used for the development of the U-Net model was increased by 10 times based on some preliminary evaluations, obtaining a total of 231 images from the original 21 images in the TR-image dataset. For each original image, the same augmentation techniques were also applied to the corresponding masks, identifying the pixels labelled as vegetal background or BMSB. In this manner, it was possible to associate each augmented image with the respective masks.

The dataset of augmented images was used to calculate a classification model using U-Net algorithm. A typical U-Net (**Figure 2**) architecture is built up from several convolutional neural layers that reduce spatial dimensionality using MaxPool layers (Ronneberger et al., 2015) in a first stage, while in the second stage increase spatial dimensionality again using either Deconvolution or Upsampling layers (Hasan et al., 2019) The U-Net's ability to transfer the entire feature map to the second stage, allows the use of higher resolution details in the later decision layers.

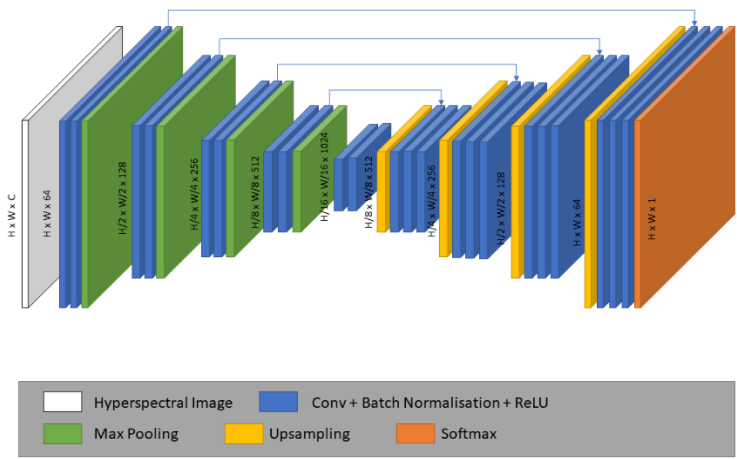

*Figure 2. Schematic representation of the U-Net architecture used in this work, where H x W x C are respectively the image height, width and number of hyperspectral channels. The color legend describes the different kind of layer used in this architecture.*

In this work, Upsampling layers were used instead of the Deconvolution layers to retain the original spatial resolution of the image. Moreover, to deal with the hyperspectral images, the first convolutional neural layer, which normally has three input dimensions (RGB channels), was adapted to the input dimensionality of the HSI image (137 spectral channels). Although this step increased the size of the network, its impact on model complexity is negligible, since later feature layers in the network contain more weights. This is clearly shown in (**Figure 2**), where the first convolution layer would normally contain 3×3×C×64 weights, where C increases from 3 dimensions for RGB to 137 dimensions for full hyperspectral. However, the tensor of the additional weights in the subsequent encoding layers are much large than the weight tensor of the first convolution layer.

In order to deal with the class unbalance between a greater amount of vegetal background pixels and the annotated BMSB pixels, a balanced loss function was introduced. This procedure allowed to give more weight to the BMSB class which would otherwise be under-represented in the loss function. Empirically, it was found that an exact reweighting based on the number of annotated BMSB and background pixels in the training set resulted in oversized BMSB detected regions due to the assignment of the border pixels between BMSB and background. Accurately sized BMSB regions were obtained by reducing the importance of annotated BMSB pixels (empirically, the annotated BMSB pixel weight was divided by 4).

The performance of the classification model was then evaluated using the same external test set images (TS-images) described in **Section 2.3.1**, corresponding to images containing different BMSB specimens from those contained in the TR-image dataset.

### 2.3.3 Classification based on merged methods

The two different classification strategies mentioned in **Section 2.3.1** and **Section 2.3.2** were merged in order to implement a more effective detection method. In particular, the reduced subset of relevant spectral bands for BMSB detection identified using s-Soft PLS-DA was used to further improve the detection ability of U-Net models. In particular, from the analysis of the s-Soft PLS-DA results, two sets of spectral bands were selected: a more restricted one, named Selection 1, and a wider one, named Selection 2. U-Net was then tested on both the spectral ranges, after adapting the convolutional neural layer to the input dimensionality of the two sets of spectral bands.

### 2.3.4 Object-level classification

In practical applications of BMSB detection, the prediction images obtained by classifying each pixel of the considered hyperspectral image can be subjected to further elaboration to obtain the final classification output at the object-level. In the present case, once the pixel-level classification models have been obtained, a further evaluation was made by focusing on the presence of clusters of neighbouring pixels predicted as BMSB. At first, defining the expected size in terms of number of pixels of BMSB specimens, it was possible to remove from the prediction images clusters of pixels predicted as BMSB with size lower than the established threshold. Then, each retained cluster of BMSB pixels was identified as a single bug.

In this study, the prediction images of test samples obtained from the different classification models (i.e., Soft PLS-DA, s-Soft PLS-DA, U-Net on full spectral range, U-Net on Selection 1 and U-Net on Selection 2) were subjected to object-level classification to simulate a real application scenario. Firstly, the clusters of neighbouring pixels predicted as BMSB with size smaller than 50 pixels were ignored. Then, the masks obtained for each image using PCA to identify the pixels belonging to the bugs were used as ground truth. The overlap between the BMSB ground truth cluster of pixels was compared with the clusters of pixels predicted as BMSB in the prediction images by computing the Intersection over Union (IoU) or Jaccard index (Jaccard, 1901). The BMSB predicted cluster of pixels was considered to match with IoU > 0.25, a slightly lower number than the standard IoU of 0.5, because the border between BMSB and background was not always clear in the hyperspectral images due to the legs of the BMSB.

In this manner, for each classification model it was possible to estimate the number of correctly identified BMSB samples (true positives), the number of actual BMSB samples not identified (false negatives), and the number of background pixels identified as BMSB (false positives).

### 2.3.5 Evaluation of classification performances

The performances of the classification models were evaluated both at the pixel-level and at the object-level. The statistical parameters used to evaluate the classification performances (Ballabio et al., 2018), considering always the BMSB class, are the following ones:

- *Sensitivity* (SENS), also referred to as *Recall* or *True Positive Rate*, defined as the ratio between the true positives (TP), i.e., the objects correctly assigned to the modelled class, and all the objects actually belonging to the considered class, i.e., the true positives and the false negatives (FN): $SENS = TP / (TP + FN)$;

- *Specificity* (SPEC), also referred to as *Selectivity* or *True Negative Rate*, defined as the ratio between the true negatives (TN), i.e., the objects correctly rejected by the modelled class, and all the objects not belonging to the considered class, i.e., the true negatives and the false positives (FP): SPEC = TN / (TN + FP). This parameter was calculated only for the pixel-based classification models, since the object-based classification only considers the BMSB samples, thus TN cannot be defined;
- *Efficiency* (EFF), defined as the geometric mean of SENS and SPEC;
- *Precision* (PREC), defined as the ratio between the true positives and all the objects assigned to the modelled class: Precision = TP / (TP + FP);
- *F1 score*, defined as the harmonic mean of SENS and PREC, is calculated as follows:

$$F1 = \frac{2}{\frac{1}{SENS}+\frac{1}{PREC}} \quad (1)$$

For the Soft PLS-DA and the s-Soft PLS-DA models all the statistical parameters were calculated at the pixel-level both in cross-validation (CV) and in prediction of the external test set of spectra (TS-spectra), and for all the classification models the relevant statistical parameters were estimated on the whole set of test images (TS-images), both at the pixel-level and at the object-level.

## 3 Results

### 3.1 Classification based on spectral features selection

**Table 1** reports the results obtained in cross-validation (CV) and prediction of the test set (TS-spectra) from the Soft PLS-DA and s-Soft PLS-DA models. For each model, the classification performances were evaluated considering SENS, SPEC, EFF, PREC and F1 score values for BMSB class. Different spectral preprocessing methods were considered to compare both the classification performances and the selected spectral regions.

|  |  | SNV + mc | | 1st derivative + mc | | Detrend + mc | | 2nd derivative + mc | |
|---|---|---|---|---|---|---|---|---|---|
|  |  | CV | TS-spectra | CV | TS-spectra | CV | TS-spectra | CV | TS-spectra |
| **Soft PLS-DA 137 spectral variables** | **LVs** | 3 | | 4 | | 5 | | 6 | |
|  | **SENS (%)** | 93.4 | 95.4 | 90.9 | 92.4 | 90.8 | 91.5 | 92.0 | 92.4 |
|  | **SPEC (%)** | 95.8 | 96.6 | 94.8 | 95.3 | 96.1 | 96.4 | 97.6 | 97.8 |
|  | **EFF (%)** | 94.6 | 96.0 | 92.8 | 93.8 | 93.4 | 93.9 | 94.8 | 95.0 |
|  | **PREC (%)** | 95.7 | 96.5 | 94.6 | 95.1 | 95.9 | 96.2 | 97.4 | 97.6 |
|  | **F1 score (%)** | 94.6 | 96.0 | 92.7 | 93.8 | 93.4 | 93.9 | 95.3 | 95.4 |
| **s-Soft PLS-DA** | **LVs** | 3 | | 5 | | 7 | | 7 | |
|  | **Selected variables** | 60 | | 73 | | 123 | | 101 | |
|  | **SENS (%)** | 92.9 | 94.9 | 91.5 | 92.5 | 92.8 | 93.0 | 92.2 | 93.0 |
|  | **SPEC (%)** | 96.0 | 96.6 | 96.4 | 96.0 | 96.0 | 96.3 | 97.6 | 97.4 |
|  | **EFF (%)** | 94.5 | 95.7 | 93.9 | 94.2 | 94.4 | 94.6 | 94.8 | 95.2 |
|  | **PREC (%)** | 95.9 | 96.5 | 96.2 | 95.9 | 95.9 | 96.1 | 97.4 | 97.2 |
|  | **F1 score (%)** | 94.4 | 95.9 | 93.9 | 94.2 | 94.3 | 94.5 | 95.3 | 95.3 |

*Table 1. Pixel-level classification results obtained by applying Soft PLS-DA and s-Soft PLS-DA in cross-validation (CV) and prediction of the test set data matrix (TS-spectra) considering different preprocessing methods.*

Generally, promising results were obtained in the discrimination between BMSB and all the considered vegetal backgrounds in cross-validation and prediction of the external test set. In addition, sparse variable selection considerably reduced the number of retained spectral variables, while maintaining satisfactory classification performances compared to the full spectral range.

Considering s-Soft PLS-DA method, second derivative row-preprocessing and mean center (mc) provided the highest cross-validation efficiency value, however the model calculated with SNV and mean center led to comparable classification performances considering at the same time a lower number of LVs and of selected spectral variables. Therefore, this model was chosen as the optimal classification model, since it is the one leading to best results in terms of both parsimony and classification performances. In addition, SNV row-preprocessing allows a simpler interpretation of the relevant spectral regions compared to second derivative preprocessing.

**Figure 3** reports the regression vector of the best s-Soft PLS-DA model (i.e., the model calculated considering SNV + mean center as spectral preprocessing method) in order to evaluate the spectral regions selected by the algorithm to discriminate BMSB from the vegetal backgrounds. It is possible to observe three main spectral bands with high absolute values in the regression vector (highlighted in green colour in **Figure 3**), and therefore with high relevance to the model. These regions correspond to the following intervals: 1220-1295 nm (C-H combination band), 1370-1410 nm (CH-combination band and O-H first overtone) and 1420-1480 nm (O-H first overtone, C=O stretch third overtone and N-H stretch first overtone). In the following, these three spectral regions will be referred to as "Selection 1".

Furthermore, two additional spectral regions were selected by the algorithm, even if these regions have low relevance to the model (highlighted in yellow color in **Figure 3**), i.e., they have low absolute values in the regression vector. These regions fall in the 980-1070 nm and 1330-1350 nm intervals, which correspond to N−H second overtone and C−H combination band, respectively. In the following, all the spectral regions selected by s-Soft PLS-DA algorithm (i.e., both the yellow and the green bars in **Figure 3**) will be referred to as "Selection 2".

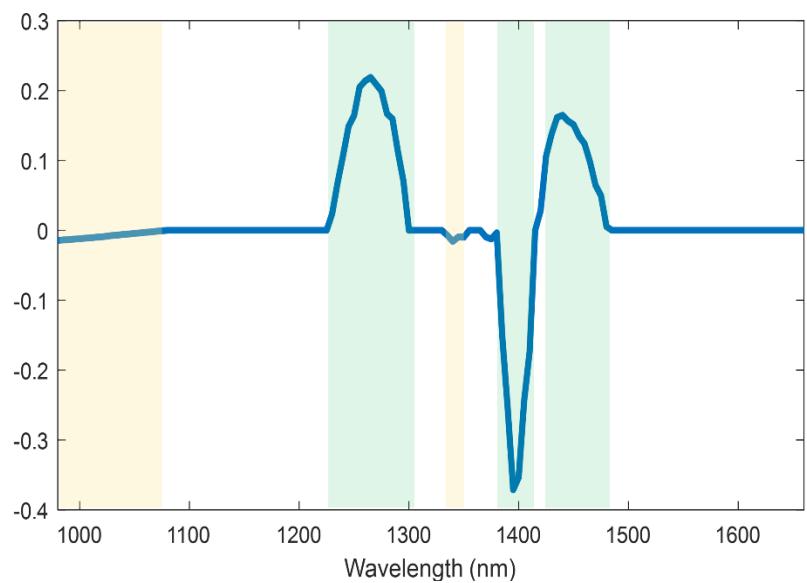

*Figure 3. Regression vector obtained by from the best s-Soft PLS-DA model and spectral regions defined as relevant for BMSB discrimination from vegetal backgrounds.*

The selected spectral regions falling in the intervals at 1220-1295 nm and at 1420-1480 nm can be associated to the presence of cellulose, hemicellulose and lignin in the different vegetal background types (Bruno & Svoronos, 2005, Jin et al., 2017, Li et al., 2015).

Conversely, the selected spectral bands falling in the intervals at 980-1070 nm, 1330-1350 nm and 1370-1400 nm correspond to absorption bands ascribable to water, protein, chitin and lipids. Therefore, these spectral regions can be associated to the biochemical structure of the outer layer of insects' exoskeleton, which is rich in chitin protein chains and lipids (Dowell et al., 1999, Johnson et al., 2020, Ridgway & Chambers, 1996).

As an additional external validation to further verify the effectiveness of the classification models, Soft PLS-DA and s-Soft PLS-DA models were applied to the whole hyperspectral images containing the bugs belonging to G4 and G5 groups (TS-images). The results obtained by applying the Soft PLS-DA and the s-Soft PLD-DA models calculated using SNV + mean center, reported in the first two columns of **Table 2**, show that s-Soft PLS-DA always leads to similar but better performances with respect to Soft PLS-DA, suggesting that variable selection provides a slight improvement of the predictive ability. The values of SENS, SPEC and EFF are always very high, while PREC, though acceptable, is much lower due to a relatively higher number of false positives with respect to the number of true positives.

| Algorithm(s) | Soft PLS-DA | s-Soft PLS-DA | U-Net | s-Soft PLS-DA + U-Net | |
|---|---|---|---|---|---|
| Spectral Variables | Full Spectrum | Selection 2 | Full Spectrum | Selection 1 | Selection 2 |
| SENS (%) | 97.1 | 97.3 | 89.2 | 79.1 | 92.4 |
| SPEC (%) | 98.4 | 98.5 | 99.2 | 96.7 | 98.8 |
| EFF (%) | 97.7 | 97.9 | 94.1 | 87.5 | 95.6 |
| PREC (%) | 66.1 | 68.3 | 79.3 | 44.1 | 71.6 |
| F1 score (%) | 78.7 | 80.2 | 83.9 | 56.6 | 80.7 |

*Table 2. Pixel-level classification results of the spectral-based (Soft PLS-DA and s-Soft PLS-DA), spatial-based (U-Net) and spectral- & spatial-based (s-Soft PLS-DA + U-Net) classification models applied to the test images.*

In more detail, by comparing the SPEC values of the TS-images predictions it is possible to point out some differences between background types (**Table S1** of Supplementary Material). In particular, slightly lower SPEC values were obtained with soil and tree branches as background types, i.e., the pixels belonging to these background types were more likely to be misclassified as BMSB. In addition, soil and tree branches were also the background types with a higher amount of not assigned pixels. However, it has to be considered that all the obtained SPEC values were higher than 95%, suggesting overall satisfactory classification results.

In **Figure 4** the prediction images of some sample images are reported, together with the corresponding RGB images as reference. Generally, from the comparison between prediction images and the corresponding RGB images, it is possible to verify that the pixels are correctly classified into the corresponding class. Furthermore, **Figure 4** highlights how difficult it is to detect BMSB specimens on dark brown vegetal backgrounds using only RGB images. Conversely, the bugs are clearly identified using NIR-HSI.

As reported in white circles in **Figure 4**, by comparing the prediction images obtained using Soft PLS-DA and s-Soft PLS-DA models considering tree branches and soil as background, it is possible to observe that spectral variable selection slightly improves the classification of background pixels. Indeed, the amount of misclassified background pixels (i.e., pixels belonging to the background that are wrongly classified as BMSB) is lower in the prediction images obtained from s-Soft PLS-DA model. Therefore, thanks to the possibility of selecting and considering only the spectral bands relevant for the classification, s-Soft PLS-DA facilitates the discrimination of background pixels, particularly when considering the background types more prone to misclassifications, like tree branches and soil.

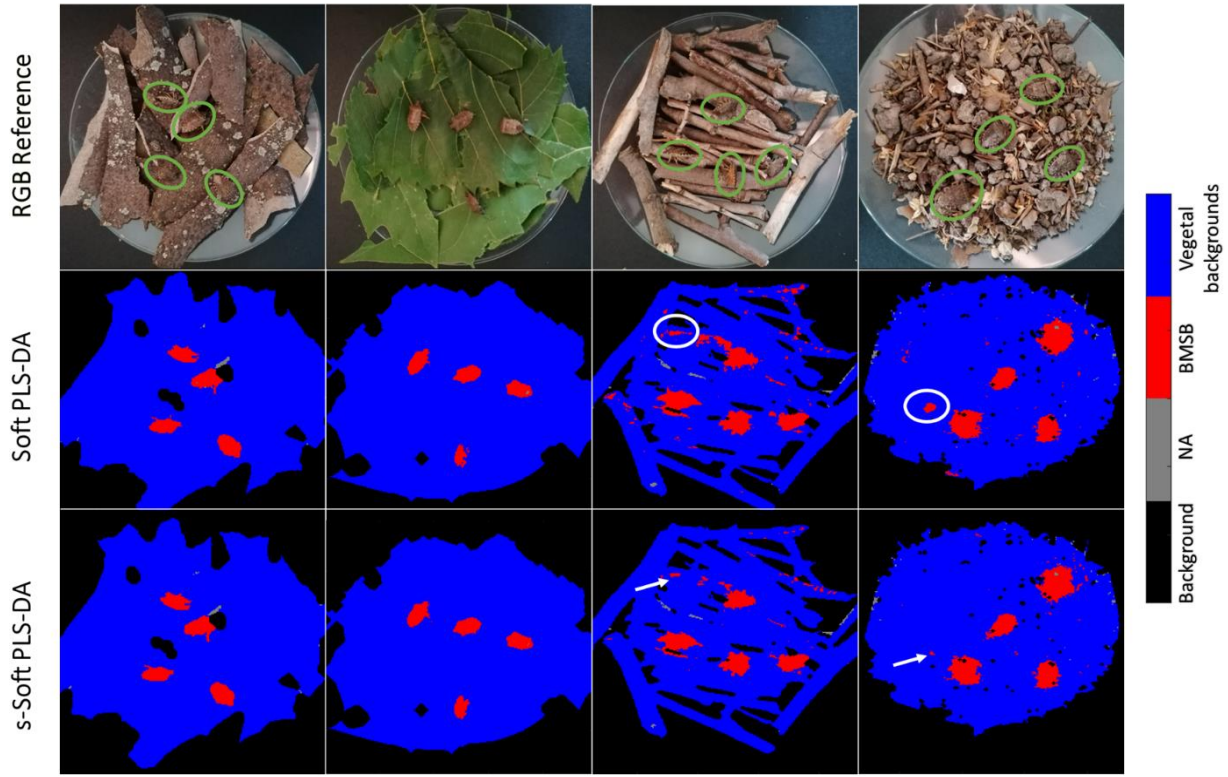

*Figure 4. Prediction images obtained by applying Soft PLS-DA and s-Soft PLS-DA models together with the corresponding RGB images of the samples.*

## 3.2 Classification based on spatial features

The third column of **Table 2** shows the SENS, SPEC, EFF, PREC and F1 score values calculated at the pixel level by applying the U-Net algorithm to the TS-images and considering the full wavelength range.

U-Net provided good classification performances, with an EFF value equal to 94.1%. Compared to Soft PLS-DA and s-Soft PLS-DA, U-Net led to a higher SPEC value; in more detail, U-Net led to higher values for 5 out of the 7 background types (**Table S2** in Supplementary Material), suggesting that this algorithm is less prone to provide false positive pixels. This aspect is also confirmed by the much higher PREC value of U-Net (79.3%) with respect to Soft PLS-DA (66.1%) and s-Soft PLS-DA (68.3%). On the other hand, the SENS value obtained with U-Net (89.2%) was much lower than the SENS values obtained with Soft PLS-DA (97.1%) and with s-Soft PLS-DA (97.3%). This fact can be attributed to misclassifications involving the pixels at the borders between BMSB and background; however, as it will be shown in Section 3.4, it did not lead to negative effects on the number of correctly detected samples. Overall, the F1 score value, accounting for both SENS and PREC, showed the highest value for U-Net (83.9%).

**Figure 5** reports the prediction images obtained by applying U-Net to the TS-images. Comparing the U-Net prediction images to those obtained from Soft PLS-DA (**Figure 4**), it can be noticed that there are fewer misclassified background pixels and that the prediction masks consist of clear clusters of pixels predicted as BMSB. However, the outlines of the BMSB samples look less detailed than those resulting from Soft PLS-DA, since U-Net also takes the neighbourhood of the pixels into account.

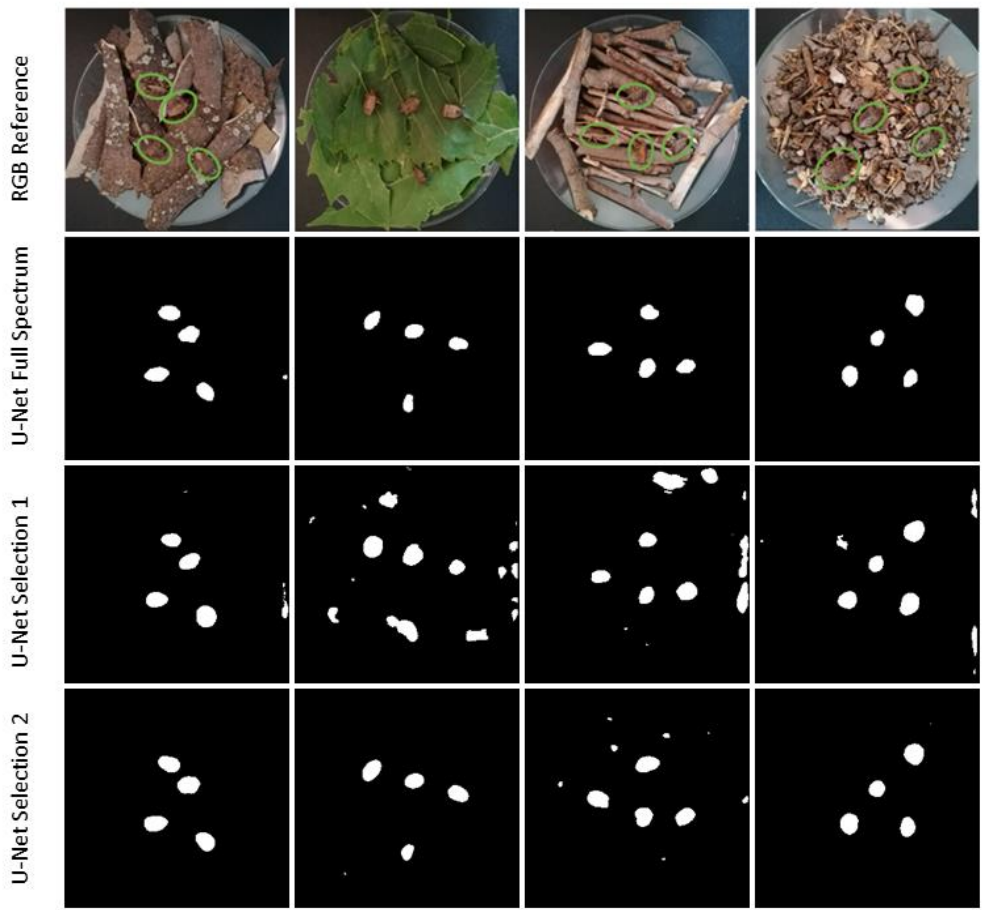

*Figure 5. Prediction masks obtained by different U-Net models together with the corresponding RGB images.*

## 3.3 Results based on merged strategies

In order to merge the strengths of linear classification methods with deep learning approach, the U-Net algorithm was applied to identify BMSB specimens considering only the wavelengths selected by s-Soft PLS-DA.

As discussed in **Section 3.1**, the linear classification algorithm s-Soft PLS-DA allowed to select 60 spectral variables out of the original 137 wavelengths, which resulted to provide important information for the discrimination between BMSB and the different vegetal backgrounds. In addition, observing the regression vector of the s-Soft PLS-DA model it was possible to identify three main spectral regions with higher relevance to the classification model. The wavelengths falling in these three more relevant spectral regions are referred to as Selection 1, while all the wavelengths selected by s-Soft PLS-DA are referred to as Selection 2.

The fourth and fifth columns of **Table 2** report the U-Net classification performances of the test set images pertaining to Selection 1 and Selection 2, respectively.

Considering the results of Selection 1, the classification performances in prediction are lower than those obtained both by the spectral-based methods and by U-Net applied to the full spectral range. In particular, the decrease in the performances is more evident considering the SENS and the PREC values (79.07% and 44.1%, respectively). These results clearly indicate that considering only the spectral regions with highest absolute values of the regression coefficients of the s-Soft PLS-DA model (green bars in **Figure 3**) is not sufficient to correctly identify the BMSB samples.

As a matter of fact, the results obtained considering Selection 2 constitute an optimal compromise between those obtained using the spectral-based methods (Soft PLS-DA and s-Soft PLS-DA) and those obtained using the spatial-based U-Net method applied to the full spectrum. Actually, merging s-Soft PLS-DA with U-Net led to a higher SENS value (92.4%) with respect to U-Net alone (89.2%), and to a higher PREC value (71.6%) with respect to Soft PLS-DA and s-Soft PLS-DA (66.1% and 66.3%, respectively). Therefore, the 60 spectral variables selected by s-Soft PLS-DA are sufficient to achieve good classification performances, with satisfactory and balanced values for all the considered statistics.

The differences in the classification performances obtained including only Selection 1 and Selection 2 spectral regions can be explained considering the absorption bands falling in the different selected intervals. Indeed, only Selection 2 includes the spectral regions falling in the 980-1070 nm and 1330-1350 nm intervals, corresponding to the absorption bands of protein, chitin and lipids, which are the main constituents BMSB exoskeleton. Therefore, these spectral bands resulted to be fundamental in discriminating BMSB from vegetal backgrounds.

**Figure 5** shows the prediction images obtained by U-net using the Full Spectrum, Selection 1 and Selection 2. The Full spectrum shows the best results followed by Selection 2, which has more false positives on wooden branches, which may be easily filtered out based on size. Selection 1 results in reduced performance due to problems with both backgrounds of leaves and wooden branches, producing false positives that are impossible to filter out.

### 3.4 Object-level evaluation of the classification models

**Table 3** reports the performance measures of the object-level classification, where only SENS, PREC and F1 score were considered for the reason discussed in **Section 2.3.5**. Notice that **Figure 5** shows the results before potential removing the smaller cluster, however the results in **Table 3** are based on the threshold mentioned in **Section 2.3.4**.

| Algorithm(s) | Soft PLS-DA | s-Soft PLS-DA | U-Net | s-Soft PLS-DA + U-Net | |
|---|---|---|---|---|---|
| **Spectral variables** | Full Spectrum | Selection 2 | Full Spectrum | Selection 1 | Selection 2 |
| **SENS (%)** | 98.2 | 98.2 | 100.0 | 82.1 | 100.0 |
| **Precision (%)** | 94.8 | 100.0 | 100.0 | 35.4 | 98.2 |
| **F1 score (%)** | 96.5 | 99.1 | 100.0 | 49.5 | 99.1 |

*Table 3. Object-level classification results of the spectral-based (Soft PLS-DA and s-Soft PLS-DA), spatial-based (U-Net) and spectral- & spatial-based (s-Soft PLS-DA + U-Net) classification models.*

Except for the U-Net model based on the spectral bands of Selection 1, all methods showed excellent results considering all three performance indicators. In particular, the U-Net model calculated considering Selection 1 showed a very low precision value, due to the presence of many clusters of background pixels wrongly classified as BMSB. As previously stated in Section 3.3, the lower performances obtained considering Selection 1 are due to the fact that the spectral regions considered in Selection 1 do not include spectral bands accounting for the main components of BMSB exoskeleton, which are instead included in Selection 2.

Comparing object-level classification performances of Soft PLS-DA and U-Net, we find that both approaches maintain high SENS values when restricting the considered wavelengths from the full wavelength range to Selection 2; in particular U-Net correctly identified all the BMSB specimens in the images in both cases. On the other hand, variable selection allowed to improve the precision value for the linear classification models (Soft PLS-DA compared to s-Soft PLS-DA or Selection 2), while the precision value slightly decreased for U-Net when considering the wavelengths of Selection 2, since in this case one false positive was identified. However, this slight decrease of performance with respect to that calculated using the whole spectral range is largely compensated by the benefits that could be obtained in practical terms. In fact, the use of a limited number of spectral bands can be the starting point for the development of classification models based on multispectral imaging systems, which require much cheaper and lighter devices than hyperspectral imaging systems.

# 4  Conclusions

NIR-HSI is a useful supporting tool for agronomists and farmers for the field monitoring of insect pests. In fact, real-time imaging techniques allow to detect consistent and sudden increase of insect populations which usually denotes an ongoing infestation. As a matter of fact, automated monitoring methods may ease the decision-making process since they permit to gather information of population dynamics and their associated ecological factors in order to develop a targeted pest control strategy. In particular, based on the outcomes of automated monitoring systems based on NIR spectral imaging, it is possible to identify specific crop areas which are more likely subjected to an ongoing infestation and require a direct inspection by technicians as well as proper pest control actions.

The present study aimed at performing a preliminary evaluation of the potential of NIR-HSI as a monitoring technique for BMSB detection.

The acquired hyperspectral images were used to develop classification models able to discriminate bugs from vegetal backgrounds following different ML strategies, simulating an in-field application scene. More in detail, the classification models were calculated considering both a linear classification algorithm (Soft PLS-DA) also combined with sparse variable selection (s-Soft PLS-DA), and a deep learning approach (U-Net). While the considered linear classification methods are based on modelling the differences of the spectral response between BMSB and vegetal backgrounds, the U-Net deep learning architecture considers non-linear spatial relationships between pixels in order to provide the classification output.

Linear classification models have the great advantages of requiring a much faster training and providing easily interpretable models, which is particularly important when dealing with spectroscopic data, since it allows to identify the most relevant spectral variables for the problem at hand. On the other hand, deep learning strategies allow to face complex classification problems when the relationship between the modelled data and the final output is not linear, but in this case the interpretation of the models is quite difficult.

In this study, both linear and non-linear approaches led alone to promising results in BMSB detection, but the most relevant outcome of this work consisted in the fact that merging these two strategies allowed to combine the strengths of both methods. In particular, spectral variable selection by s-Soft PLS-DA was used in order to select a subset of relevant variables to be used for the classification with U-Net, leading to classification performances comparable to those obtained by U-Net for the full wavelength range.

The results obtained in this study can be considered as a first step toward the development of multispectral imaging systems for the detection of BMSB. Indeed, MSI systems are more suitable for applications in the field thanks to their relatively low costs and higher resistance of the optical components. Future work will focus on implementing the spectral regions selected by s-Soft PLS-DA into a MSI-based monitoring system and on evaluating the effectiveness of the proposed approach on real field conditions. To have a preliminary assessment of the performances that can be reached with the multispectral system, hyperspectral data can be used to simulate a multispectral imaging system embedding only band-pass filters falling in the selected spectral regions (Calvini et al., 2017).

Furthermore, it will also be necessary to face the issues that may arise from practical applications of multispectral systems in field. One problem consists in the fact that the images can be acquired at different distances from the camera, resulting in images with different resolutions. However, this problem can be easily tackled by creating augmented images of the insects at different scaling levels.

Another problem that has to be considered in practical applications is the presence of insects different from BMSB. Future studies will consider the possibility of identifying different insect species by combining spectral information with spatial and morphological features resulting from the object-level classification.

## Acknowledgements

Authors wish to thank HALY.ID, project of ERA-NET Cofund ICT-AGRI-FOOD, with funding provided by national sources (Ministero delle politiche agricole e forestali, MIPAAF) and co-funding by the European Union's Horizon 2020 research and innovation program, Grant Agreement number 862671.

Rosalba Calvini would like to thank the Italian funding programme *Fondo Sociale Europeo REACT-EU - PON "Ricerca e Innovazione" 2014 – 2020 – Azione IV.6 Contratti di ricerca su tematiche Green (D.M. 1062 del 10/08/ 2021)* for supporting her research (CUP: E95F21002330001; contract number 17-G-13884-4).

OnePlanet Research Center is supported by the Province of Gelderland.

# Author contributions

V. Ferrari – Methodology; Software; Formal analysis; Investigation; Data curation; Writing – Original Draft; Writing – Review & Editing

R. Calvini – Conceptualization; Methodology; Software; Data curation; Writing – Original Draft; Writing – Review & Editing

B. Boom –Methodology; Software; Formal analysis; Investigation; Data Curation; Writing – Review & Editing

C. Menozzi – Investigation; Writing – Review & Editing

A. K. Rangarajan – Investigation; Writing – Review & Editing

P. Offermans – Conceptualization; Methodology; Writing – Review & Editing; Supervision; Project administration; Funding acquisition

L. Maistrello –Methodology; Resources; Writing – Review & Editing; Supervision; Project administration; Funding acquisition

A. Ulrici – Conceptualization; Methodology; Writing – Review & Editing; Supervision; Project administration; Funding acquisition

# Supplementary Material

| | Linear Classification Models Outputs | | | | | | | | | | | | | |
|---|---|---|---|---|---|---|---|---|---|---|---|---|---|---|
| | Bark | | Grass | | Dry leaves | | Green leaves | | Yellow leaves | | Soil | | Tree branches | |
| | Soft PLSDA | s-Soft PLSDA | Soft PLSDA | s-Soft PLSDA | Soft PLSDA | s-Soft PLSDA | Soft PLSDA | s-Soft PLSDA | Soft PLSDA | s-Soft PLSDA | Soft PLSDA | s-Soft PLSDA | Soft PLSDA | s-Soft PLSDA |
| SENS (%) | 98.0 | 98.0 | 97.9 | 98.6 | 89.1 | 88.2 | 95.8 | 97.3 | 99.9 | 99.9 | 99.4 | 99.4 | 99.7 | 99.4 |
| SPEC (%) | 99.4 | 99.4 | 98.3 | 97.9 | 99.4 | 99.4 | 99.5 | 99.4 | 98.6 | 98.4 | 97.1 | 97.2 | 95.4 | 97.0 |
| EFF (%) | **98.7** | **98.7** | **98.1** | **98.3** | **94.1** | **93.6** | **97.6** | **98.4** | **99.2** | **99.2** | **98.2** | **98.3** | **97.5** | **98.2** |
| BMSB N/A (%) | 0.6 | 0.5 | 0.2 | 0.3 | 0.0 | 0.0 | 0.3 | 0.3 | 0.0 | 0.1 | 0.4 | 0.4 | 0.1 | 0.0 |
| BACK N/A (%) | 0.2 | 0.2 | 0.0 | 0.0 | 0.1 | 0.1 | 0.0 | 0.0 | 0.0 | 0.0 | 0.3 | 0.23 | 0.5 | 0.4 |

**Table S1** Comparison between linear classification models: prediction performances for TS-images. For each background are reported the percent values of SENS, SPEC, EFF for the BMSB class, and not assigned (N/A) pixels for both BMSB and background classes, obtained considering the full wavelength range (Soft PLS-DA) and the spectral regions selected by s-Soft PLS-DA.

|  | Full wavelength range | | | | | | | | | | | | | |
|---|---|---|---|---|---|---|---|---|---|---|---|---|---|---|
|  | Bark | | Grass | | Dry leaves | | Green leaves | | Yellow leaves | | Soil | | Tree branches | |
|  | Soft PLS-DA | UNET | Soft PLS-DA | UNET | Soft PLS-DA | UNET | Soft PLS-DA | UNET | Soft PLS-DA | UNET | Soft PLS-DA | UNET | Soft PLS-DA | UNET |
| SENS (%) | 98.0 | 92.1 | 97.9 | 79.1.8 | 89.1 | 95.8 | 95.8 | 88.0 | 99.9 | 97.2 | 99.4 | 84.0 | 99.7 | 90.6 |
| SPEC (%) | 99.4 | 99.0 | 98.3 | 99.4 | 99.4 | 98.3 | 99.5 | 99.7 | 98.6 | 99.1 | 97.1 | 99.6 | 95.4 | 99.3 |
| EFF (%) | 98.7 | 95.5 | 98.1 | 88.7 | 94.1 | 97.0 | 97.6 | 93.7 | 99.2 | 98.2 | 98.2 | 91.5 | 97.5 | 94.9 |
| PREC (%) | 83.5 | 73.9 | 71.1 | 86.0 | 82.9 | 66.5 | 83.6 | 88.9 | 61.8 | 71.8 | 57.5 | 89.3 | 47.5 | 84.6 |
| F1 score (%) | 90.2 | 82.0 | 82.4 | 82.4 | 85.9 | 78.5 | 89.3 | 88.4 | 76.4 | 82.6 | 72.8 | 86.6 | 64.3 | 87.5 |

**Table S2** Comparison between linear and non-linear classification methods: prediction performances for TS-images. For each background are reported the percent values of SENS, SPEC, EFF, PREC and F1 score obtained by considering the full wavelength range.

|  | Spectral regions selected (Selection 2) | | | | | | | | | | | | | |
| --- | --- | --- | --- | --- | --- | --- | --- | --- | --- | --- | --- | --- | --- | --- |
|  | Bark | | Grass | | Dry leaves | | Green leaves | | Yellow leaves | | Soil | | Tree branches | |
|  | s-Soft PLS-DA | UNET | s-Soft PLS-DA | UNET | s-Soft PLS-DA | UNET | s-Soft PLS-DA | UNET | s-Soft PLS-DA | UNET | s-Soft PLS-DA | UNET | s-Soft PLS-DA | UNET |
| SENS (%) | 98.0 | 93.0 | 98.6 | 91.2 | 88.2 | 92.2 | 97.3 | 93.8 | 99.9 | 96.9 | 99.4 | 91.1 | 99.4 | 90.3 |
| SPEC (%) | 99.4 | 98.6 | 97.9 | 98.5 | 99.4 | 98.6 | 99.4 | 99.3 | 98.4 | 98.7 | 97.2 | 99.1 | 97.0 | 98.4 |
| EFF (%) | 98.7 | 95.7 | 98.3 | 94.8 | 93.6 | 95.4 | 98.4 | 96.5 | 99.2 | 97.8 | 98.3 | 95.0 | 98.2 | 94.3 |
| PREC (%) | 82.5 | 66.3 | 67.3 | 72.1 | 84.8 | 70.4 | 81.3 | 78.5 | 59.6 | 63.4 | 59.0 | 81.1 | 58.0 | 70.3 |
| F1 score (%) | 89.6 | 77.4 | 80.0 | 80.5 | 86.5 | 79.9 | 88.6 | 85.5 | 74.7 | 76.7 | 74.1 | 85.8 | 73.3 | 79.0 |

**Table S3** Comparison between linear and non-linear classification methods: prediction performances for TS-images. For each background are reported the percent values of SENS, SPEC, EFF, PREC and F1 score obtained by considering the Selection 2 spectral region.